\RequirePackage{fix-cm}
\documentclass[smallextended]{svjour3}       
\smartqed  
\usepackage{graphicx}
\usepackage{marvosym}
\usepackage [latin1]{inputenc}
\begin{document}

\title{The phase matching quantum key distribution protocol with 3-state systems
}


\author{Han Duo\and
        Li Zhihui\textsuperscript{\Letter}\and
        Liu Chengji\and
        Gao Feifei 
}


\institute{Han Duo \at
              School of Mathematics and Information Science, Shaanxi Normal University, Xi'an, China, 710119. \\
              Tel.: +18829287176\\
              \email{handd@snnu.edu.cn}           
           \and
           Li Zhihui\textsuperscript{\Letter}\at
              School of Mathematics and Information Science, Shaanxi Normal University, Xi'an, China, 710119. \\
              Tel.: +13032989886\\
              \email{lizhihui@snnu.edu.cn}
              \and
           Liu Chengji \at
              State Key Laboratory of Integrated Services Networks, Xidian University, Xian, China, 710071.\\
              Tel.: +15091056770\\
              \email{120148147@qq.com}
              \and
           Gao Feifei \at
              School of Mathematics and Information Science, Shaanxi Normal University, Xi'an, China, 710119. \\
              \email{7920999462@qq.com}
}

\date{Received: date / Accepted: date}

\maketitle

\begin{abstract}
Quantum Key Distribution, as a branch of quantum mechanics in cryptography, can distribute keys between legal communication parties in an unconditionally secure manner, thus can realize in transmitting confidential information with unconditional security. We consider a Phase-Matching Quantum Key Distribution protocol with 3-state systems for the first time, where the phase of the coherent state is 3,thus we propose three different ways to response to every successful detection and two parties gain their raw keys by ``flip and flip". The simulation results show that compared with Phase-Matching Quantum Key Distribution protocol where the phase equals 2, the proposed protocol breaks the limit of linear key generation rate in a shorter distance, and the longest practical transmission distance is about 470 $km$, whereas the ones of BB84 protocol is lower than 250 $km$.
\end{abstract}

\section{Introduction}
\label{intro}
\hangafter=-1\hangindent=14pt\noindent Quantum cryptography \cite{Ref1} is an interdisciplinary subject combining cryptography and quantum mechanics \cite{Ref2} .It is an important research topic.Its security is based on the basic principles of quantum mechanics, such as quantum non-cloning theorem, uncertainty principle \cite{Ref3} et al., the quantum key distribution technology in quantum cryptography provides a means of communication for both parties to obtain unconditional security keys. Security and practical applications are the core of this research.
\par The first quantum cryptography protocol,the BB84 quantum key distribution protocol \cite{Ref1} , was proposed by Bennett and Brassard in 1984, which introduced quantum mechanics into practical applications. However, until 1999, Lo and Chau proved the security of the BB84 protocol by equating the BB84 protocol with an entanglement and purification protocol \cite{Ref4} ;but the quantum computer was needed in the proof process. Then in 2000, Shor and Preskill proposed a more concise proof method for CSS quantum error correction code for entanglement and purification \cite{Ref5} ;which removed the dependence on quantum computers. Lo et al's research further proved bit error correction and phase error correction can be implemented separately \cite{Ref6} .
\par Although the theoretical security of the BB84 protocol has been proved, its actual implementation still has a large security risk. Since there is no ideal single-photon source \cite{Ref7} in practice, a weak coherent pulse (Weak Coherent Pulse) \cite{Ref8} is commonly used to simulate a single-photon source, which leads to the generation of Photon Number Splitting \cite{Ref9} . In 2003, the problem was solved for the first time, since a decoy-state scheme \cite{Ref10} was proposed to defend against PNS attacks. Besides the hidden dangers of the light source, the detector side channel attack \cite{Ref11} also greatly threatened the security of the password. HK proposed decoy Measurement-Device-Independent Quantum Key Distribution (MDI-QKD) protocol \cite{Ref12} in 2012, which eliminates the detector side channel attack \cite{Ref13} without introducing more implementation equipment and double the transmission distance covered by the traditional QKD scheme at the same time.
\par However, these QKD protocols has same limitations---they never exceed the limit of the Secret Key Capacity (SKC) \cite{Ref14} of the lossy optical quantum channel. X.B.Wang et al. proposed the Twin-field Quantum Key Distribution (TF-QKD) protocol \cite{Ref15} in 2018, which broke the SKC bound of the previous QKD protocols under the condition of ensuring key security. The square root dependence of the key generation rate on the channel transmittance is obtained. However, the security of the agreement has not been proven. X.F.Ma et al. proposed the Phase-matching Quantum Key Distribution (PM-QKD) protocol in 2018 \cite{Ref16}  which illustrated a security proof based on optical mode, and resisted all possible measuring attacks.
\par In the PM-QKD protocol, the communication parties Alice and Bob each generate coherent state pulses independently. For a d-phase PM-QKD protocol, Alice and Bob encode their key information $\kappa_{a}, \kappa_{b} \in\{0,1, \cdots \cdot d-1\}$ into the phase of the coherent state. Paper\cite{Ref16} mainly studied the the PM-QKD in the case of phase d=2 (2-PM-QKD protocol) with phase randomization. In theory, the protocol is immune to all possible measurement attacks, and its key rate can even exceed the transmission probability $\eta$ between two communicating parties; In practice, the protocol applies phase compensation to devise a practical version of the scheme without phase locking \cite{Ref17} , which makes the proposed scheme feasible in current technology.
\par Inspired by the PM-QKD protocol in \cite{Ref18} , this paper proposes a new PM-QKD whose phase d=3 with phase randomization(For simplicity,we use the name ``3-PM-QKD protocol" in the text below).
\par In the 2-PM-QKD protocol, each transmitted 32-bit binary bit can encode the largest unsigned number, but if the 3-PM-QKD protocol is used, every 32-bit ternary trit can be successfully transmitted. In addition, in this protocol, the range of random phase matching is wider and the probability of successful matching is higher. Alice and Bob retain their key trits when their declared random phases difference is 0, $\frac{2 \pi}{3}$, or $\frac{4 \pi}{3}$, which significantly improves the success rate of the phase sifting phase and also results in a higher final key rate.
\par The paper organized as follows. Following the 2-PM-QKD protocol, we propose the 3-PM-QKD protocol that can surpass the linear key-rate bound and make the key rate increase, whose details are given In Sec.2. Then, the security of 3-PM-QKD is proved in Sec.3, and in Sec.4, we consider all practical factors to simulate the 3-PM-QKD key rate and compare it to the previous QKD protocol. Finally, we summarize this work , put forward the the $n$-PM-QKD protocol and expound its some curious features in Sec.5.

\section{3-PM-QKD protocol}
\label{sec:1}
\hangafter=-1\hangindent=14pt\noindent This paper proposes a 3-PM-QKD protocol with phase randomization.That is,Alice and Bob add extra random phases on their coherent state pulses before sending these pulses to Eve. After Eve's announcement,Alice and Bob announce the extra random phases and postselect the signals based on their random phases.The specific steps and related descriptions are as follows.
\subsection{Specific steps}
\label{sec:2}
\hangafter=-1\hangindent=14pt\noindent \textbf{Step1}  State Preparation - Alice randomly generates a key trit $\kappa_{a}\in\{0,1,2\}$and a random phase $\phi_{a} \in[0,2 \pi)$, and then prepares a coherent state $\left|\sqrt{\mu_{a}} e^{i\left(\phi_{a}+\frac{2 \pi}{3} \kappa_{a}\right)}\right|_{A}$. Similarly, Bob generates $\kappa_{b} \in\{0,1,2\}$ and $\phi_{b} \in[0,2 \pi)$ then prepare $\left|\sqrt{\mu_{b}} e^{i\left(\phi_{b}+\frac{2 \pi}{3} \kappa_{b}\right)}\right|_{B}$.

\par \textbf{Step2}  Measurements - Alice and Bob send their light pulses A and B to an untrusted Eve,which needs to perform interferometry and record the response detector ($D_{0}$, $D_{1}$, or $D_{2}$).
In particular,the detector response rules are as follows:
$$\Delta_{\phi}=\left|\phi_{a}+\frac{2 \pi}{3} \kappa_{a}-\left(\phi_{b}+\frac{2 \pi}{3} \kappa_{b}\right)\right|=\left|\frac{2 \pi}{3}\left(\kappa_{a}-\kappa_{b}\right)+\left(\phi_{a}-\phi_{b}\right)\right|.$$
\par The way the detector responds in this protocol depends on the phase difference between Alice and Bob,The detector response mechanism is set to:
$$\left\{
\begin{array}{rcl}
&&D_{0}\;response,\;when\Delta\phi=0 \; (\bmod 2 \pi),\\
&&D_{1}\;response,\;when\Delta\phi=\frac{2 \pi}{3} \; (\bmod 2 \pi),\\
&&D_{2}\;response,\;when\Delta\phi=\frac{4 \pi}{3} \; (\bmod 2 \pi).\\
\end{array} \right. $$

\par \textbf{Step3}  Statement - Eve announces his detection result. Then Alice and Bob announce random
phase $\phi_{a}$ and $\phi_{b}$ ,respectively.

\par \textbf{Step4}  Sifting - Alice and Bob repeat the above steps multiple times. When Eve announces a successful response (just one detector response),Alice and Bob make the $\kappa_{a}$ and $\kappa_{b}$ the raw key trits.

According to Eve's statement, Bob flips his key trits $\kappa_{b}$ accordingly.The flipped key trits are
recorded as $\kappa_{b}^{\prime}$. The flip rule is as follows:
$$\left\{
\begin{array}{rcl}
&&\kappa_{b}^{\prime}=\kappa_{b},\;if\; D_{0}\; responce, \\
&&\kappa_{b}^{\prime}=\kappa_{b}+1(\bmod 2 \pi),\;if\; D_{1}\; responce, \\
&&\kappa_{b}^{\prime}=\kappa_{b}+2(\bmod 2 \pi),\;if\; D_{2}\; responce.\\
\end{array} \right. $$

\par When Alice and Bob respectively announce random phases, Bob flips his key trits $\kappa_{b}^{\prime}$ again according to their random phase difference $\left|\phi_{a}-\phi_{b}\right|$ .The flipped key trits are recorded as $\kappa_{b}^{\prime \prime}$.Theflipping rules are as follows:
$$\left\{
\begin{array}{rcl}
&&\kappa_{b}^{\prime\prime}=\kappa_{b}^{\prime}(\bmod 2\pi),\;if\left|\phi_{a}-\phi_{b}\right|=0 \; (\bmod 2 \pi),\\
&&\kappa_{b}^{\prime\prime}=\kappa_{b}^{\prime}-1(\bmod 2\pi),\;if\left|\phi_{a}-\phi_{b}\right|=\frac{2 \pi}{3} \; (\bmod 2 \pi),\\
&&\kappa_{b}^{\prime\prime}=\kappa_{b}^{\prime}-2(\bmod 2\pi),\;if\left|\phi_{a}-\phi_{b}\right|=\frac{4 \pi}{3} \; (\bmod 2 \pi).\\
\end{array} \right. $$

\par Finally, Bob's key trits are $\kappa_{b}^{\prime \prime}$.
\par \textbf{Step5}  Parameter Estimation - Alice and Bob derive the gain $Q_{\mu}$ and quantum trit error rate $E_{\mu}^{Z}$ from all the retained raw data and then estimate $E_{\mu}^{X}$.

\par \textbf{Step6}  Key Distillation - Alice and Bob perform error correction and privacy amplification on the sifted key trits to generate a private key (note that the error correction and privacy amplification of this protocol are the same as in all QKD protocols except that we must use trits Not bits, so the parity becomes a ternary test, ie the modulus is 3 \cite{Ref18} ).

\par In the actual implementation of this protocol, Alice and Bob retain their signals only when their declared random phases difference is 0, $\frac{2 \pi}{3}$, or $\frac{4 \pi}{3}$. However, due to the announcement of phase continuity, the probability of successful sifting is 0. In addition, the phase locking technique required in actual implementation is very difficult . Therefore, we use the phase post compensation method \cite{Ref19} like paper.

\par The post-phase compensation method used here is similar to 2-PM-QKD protocol. Alice and Bob divide the phase interval $[0,2 \pi)$ into M slices first. When a random phase is declared, Alice and Bob only compare the slice indicators, not the exact phase. This makes the step of phase sifting practical, but introduces inherent bias errors. This bias error can compensate for the inherent bias error by sacrificing a portion of the data, minimizing the quiz error rate QBER based on random sampling, and calculating the appropriate phase offset. In addition, Alice and Bob do not perform phase sifting immediately in each round,but perform this phase sifting in data post-processing. This makes the 3-PM-QKD protocol practical.

\par More importantly,in the 2-PM-QKD protocol, each successfully transmitted 32-bit binary bit can encode the largest unsigned number to $2^{32}-1$ ,but if this protocol is used,every 32-trit ternary trit transmitted can be encoded to $3^{32}-1$ . When the transmission efficiency is same, every coherent state can carry more information.

\par In this protocol,the range of random phase matching is wider and the probability of matching success is higher. Alice and Bob retain their signals when their declared random phases difference is 0, $\frac{2 \pi}{3}$, or $\frac{4 \pi}{3}$, which significantly improves the success rate of the phase screening phase and also results in a higher final key rate.

\par To illustrate the feasibility of this protocol, this paper presents a simple key correspondence table to illustrate how Alice and Bob match key information by ``flip and flip" successfully.

\subsection{Key-correspondence table of this protocol}

In the 3-PM-QKD protocol, the key information $\kappa_{a(b)} \in\{0,1,2\}$, so a successful key is generated if and only if the random phase difference is an integer multiple $\frac{2 \pi}{3}$ (less than anon-negative integer multiple of 3).There are 27 cases, and we list the situation when $\phi_{a}$ and $\phi_{b}$ satisfy with $\left|\phi_{a}-\phi_{b}\right|=\frac{4 \pi}{3}(\bmod 2 \pi)$ in this section, and other cases are equally available.


\begin{table}[!h]
\caption{Key-correspondence table with
$\left|\phi_{a}-\phi_{b}\right|=\frac{4 \pi}{3}(\bmod 2 \pi)$}
\label{tab:1}       
\begin{tabular}{lllllll}
\hline\noalign{\smallskip}
$\kappa_{a}$&$\kappa_{b}$&$\left|\phi_{a}-\phi_{b}\right|$&$\Delta_{\phi}$&Response&$\kappa_{b}^{\prime}$&$\kappa_{b}^{\prime \prime}$  \\
\noalign{\smallskip}\hline\noalign{\smallskip}
0&0&$4 \pi / 3$&$4 \pi / 3$&$D_{2}$&2&0\\
0&1&$4 \pi / 3$&$2 \pi / 3$&$D_{1}$&2&0\\
0&2&$4 \pi / 3$&0&$D_{1}$&2&0\\
1&0&$4 \pi / 3$&0&$D_{1}$&0&1\\
1&1&$4 \pi / 3$&$4 \pi / 3$&$D_{1}$&0&1\\
1&2&$4 \pi / 3$&$2 \pi / 3$&$D_{1}$&0&1\\
2&0&$4 \pi / 3$&$2 \pi / 3$&$D_{1}$&1&2\\
2&1&$4 \pi / 3$&0&$D_{1}$&1&2\\
2&2&$4 \pi / 3$&$4 \pi / 3$&$D_{1}$&1&2\\
\noalign{\smallskip}\hline
\end{tabular}
\end{table}

The Table \uppercase\expandafter{\romannumeral1} shows that accurate detection response and key sifting can guarantee a successful key match with a probability close to 1, and the probability of successful match is higher than that of  2-PM-QKD protocol.

\section{Security of 3-PM-QKD protocol}

\hangafter=-1\hangindent=14pt\noindent Unlike the general QKD protocol security proof, the commonly used photon number channel model \cite{Ref20} and the ``tagging" method used in the security proof by Gottsman et al.(GLLP security proofs) \cite{Ref21} are no longer applicable here. This is because the random phases of Alice and Bob are annonced in this protocol, and the quantum source can no longer be regarded as a mixture of photon number states. But as mentioned in  \cite{Ref16} , we can directly analyze the optical mode by applying Lo-Chau entangled distillation theory to demonstrate the security of PM-QKD with coherent pluses.

\par The proof of the security of this protocol is followed by the analysis of distillable entanglement based on the equivalent entanglement protocol, and transforms it into 3-PM-QKD protocol gradually. The security performance is proved in each operation. The proof process is similar to the 2-PM-QKD protocol, except we must use trits instead of bits, so we will not repeat them here.

\section{Simulation Results}

\hangafter=-1\hangindent=14pt\noindent Since our solution is generalized from 2-PM-QKD, the simulation in this section is mainly compared with the 2-PM-QKD protocol. According to [16], when $l>120 k m$, the key rate of 2-PM-QKD exceeded the key rate of traditional BB84 protocol; when transmitting distance $l>250 k m$, 2-PM-QKD could exceed the limit of linear key rate; Compared with MDI-QKD, 2-PM-QKD can achieve a longer transmission distance of $l=450 \mathrm{km}$, and at the time $l>300 k m$, the key rate increased by about 4-6 orders of magnitude. This section will prove that the 3-PM-QKD protocol is better than the 2-PM-QKD protocol, and thus better than the traditional BB84 protocol and the MDI-QKD protocol.

\par Applying the key rate formula in Shor-Preskill's security proof [5]

\begin{equation}
r=1-H\left(E^{Z}\right)-H\left(E^{X}\right).
\end{equation}

\par And the key rate formula in [16]

\begin{equation}
R_{2 - P M} \geq \frac{2}{M} Q_{\mu}\left[1-f H\left(E_{\mu}^{z}\right)-H\left(E_{\mu}^{X}\right)\right].
\end{equation}

\par Where $E^{Z}$ and $E^{X}$ are the $Z$ error rate and the $X$ error rate, respectively. $H(x)=-x \log _{2} x-(1-x) \log _{2}(1-x)$ are the binary Shannon entropy function, $Q_{\mu}$ is the phase error rate, and $f$ is the error correction efficiency.

\par Because we have only expanded the value range of the key bits in this paper, other parts of the key rate formula are still similar to the 2-PM-QKD protocol, but the phase sifting factor in this article is $3 / \mathrm{M}$, Shannon's entropy function becomes $H(x)=-x \log _{3} x-(1-x) \log _{3}(1-x)$ for the three-state system.Finally, our key rate formula is

\begin{equation}
R_{3 -PM} \geq \frac{3}{M} Q_{\mu}\left[1-f H\left(E_{\mu}^{z}\right)-H\left(E_{\mu}^{X}\right)\right].
\end{equation}

\par We use the parameters given in the Table 2 below to simulate the performance of 3-PM-QKD. Assuming that the lossy channels of Alice and Bob are symmetrical; the dark count rate is from Ref. \cite{Ref22} , and other parameters are set to classic values. (Note that in order to identify the effect of the key rate is on the protocol itself rather than other parameters , the actual setting parameters used in this article are completely consistent with the 2-PM-QKD protocol).

\begin{table}[h]
\caption{Parameters used for simulation in 3-PM-QKD protocol}
\label{tab:1}       
\begin{tabular}{ll}
\hline\noalign{\smallskip}
Parameters&Values  \\
\noalign{\smallskip}\hline\noalign{\smallskip}
Dark count rate $p_{d}$&$8 \times 10^{-8}$\\
Error correction efficiency $f$&1.15\\
Detector efficiency $\eta_{d}$&14.5\%\\
Number of phese slices $M$&16\\
Misalignment error $e_{d}$&1.5\%\\
\noalign{\smallskip}\hline
\end{tabular}
\end{table}

\par The simulation results are shown in the following figure.

\begin{figure}[h]
  \includegraphics{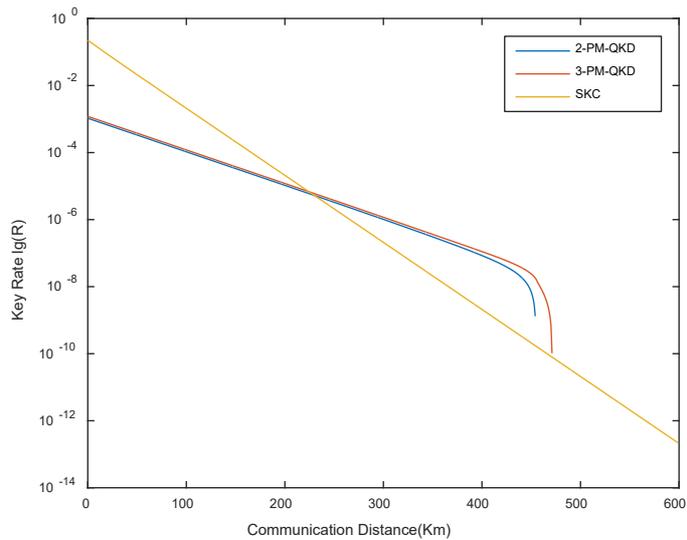}
\caption{Simulation of our protocol. For the considered simulation parameters, the key rate is similar to 2-PM-QKD protocol, but it breaks the SKC bound in a shorter distance and the effective transmission distance has increased by about 20 kilometers.}
\label{fig:1}       
\end{figure}

As can be seen, our protocol has a small increase on key rate compared with the 2-PM-QKD protocol, but the effective transmission distance has increased by about 20 kilometers, and it has broken through SKC bound in a shorter distance.

\section{Summary and Outlook}

\hangafter=-1\hangindent=14pt\noindent This paper proposes the 3-PM-QKD protocol and proves its security. The 3-PM-QKD protocol not only breaks the boundaries of SKC, but also reduces the distance to break the boundaries of SKC,meanwhile,the proposed protocol increases the key rate and effective transmission distance.

\par Also,the higher-dimensional promotion of the 3-PM-QKD protocol in this paper will be an interesting direction, such as the $n$-PM-QKD protocol with phase randomization. This is a very difficult problem, because when space is extended to any dimension, it is difficult to express all its properties strictly, but this attempt is very interesting. At the time $\kappa_{a(b)} \in\{0,1,2, \cdots, n\}$, although the selection interval of the random phase was still the same, the probability of successful screening could approach 100\%.This is an interesting change, which is because there is always a exact detector response for the specific value of any phase difference, but this requires extremely accurate detector standards. In any case, this will be one of the efforts of QKD protocol in practice.The specific steps of $n$-PM-QKD is similar to 2-PM-QKD, which will not be described in detail here.

\par However, the $n$-PM-QKD protocol is currently only possible theoretically, but if it can be successfully implemented, that will greatly increase the key rate of the QKD protocol and guarantee a 100\% probability of successful phase matching, in which the corresponding parameter estimation and security proof will be the largest challenge . Once the phase of the coherent state is extended to infinite dimensions, it may have some distinctive properties, which is a subject worthy of study.

\begin{acknowledgements}
We would like to thank anonymous reviewer for valuable comments.This work is supported by the National Natural Science Foundation of China under Grant No.11671244.
\end{acknowledgements}

%
%

\bibliographystyle{spphys}       


\end{document}